\documentclass[12pt]{iopart}
\usepackage{iopams}
\usepackage{cite}

\begin{document}
\title[A ... new approach of quantum measurements]
{A possible new approach of quantum measurements}
\author{S. Dumitru}
\address{Department  of Physics, ``Transilvania'' University,
B-dul Eroilor 29, R-2200 Brasov, Romania}
\ead{s.dumitru@unitbv.ro}
\begin{abstract}
It is proposed a possible new approach of quantum measurements
(QMS), disconnected of the traditional interpretation of
uncertainty relations and independent of any appeal to the strange
idea of collapse (reduction) of wave functions. The new approach
regards QMS as a statistical samplings (but not as simple
detection acts) and their description as a distinct task,
independent of actual procedures of quantum mechanics. A QMS is
described by means of transformations of probability density and
probability current, from intrinsic into recorded readings. The
quantum observables appear as random variables, described by usual
operators and valuable through probabilistic numerical parameters
(mean values, correlations and standard deviations). The values of
the respective parameters are not the same in the two mentioned
readings. Then the measurement uncertainties (errors) are
described by means of the changes in the alluded values. The new
QMS approach is illustrated through an one-dimensional example.
\end{abstract}
\submitto{JPA}
\maketitle

\section{Introduction}
In connection with the foundation and interpretation of quantum
mechanics (QM) the description of quantum measurements (QMS) is a
problem often considered \cite{1} as: \emph{``probably the most
important part of the theory''}. The respective problem germinate
from the discussions about the traditional interpretation of
uncertainty relations (TIUR). In its essence the mentioned problem
refers to the theoretical descriptions of measurements regarding
the observables (physical quantities) specific for quantum
microparticles. Along the years a large number of works reported
approaches of QMS problem (for a significant and updated
bibliography see \cite{1} and preprints archives \cite{2,3}). As a
notable aspect today one finds that many of the mentioned
approaches are TIUR-connected, because they are founded on
conjectures induced (inspired) someway from TIUR. In the main the
respective approaches, as well as the TIUR, are centered round the
idea that the uncertainty relations (UR) are capital physical
formulas with an exclusive quantum (i.e. non-classical)
significance.

On the other hand, if it is subjected to a minute re-examination,
TIUR proves oneself to be nothing but an incorrect doctrine that
must be denied. Such a re-examination was developed progressively
in our works \cite{4,5,6,7,8,9} and its essential conclusions, of
interest here, can be found in the recent paper \cite{10}.
Through the mentioned re-examination of TIUR one finds that UR
must be be reinterpreted in a more natural manner - i.e as
relations belonging to a more general family of formulas (from
both quantum and classical physics) which regard the fluctuations
of observables with random characteristics. Moreover UR must be
deprived of any capital (or extraordinary) attributes usually
asserted by TIUR and assumed by the TIUR-connected approaches of
QMS.

In the mentioned circumstances as regards the QMS problem becomes
of actual interest to search new approaches disconnected from
TIUR doctrine. Such an approach is the aim of the present paper.
For our aim in the next section we present the main conjectures
of the alluded TIUR-connected approaches as well as the
corresponding shortcomings. Subsequently, in Sec.3. , we present
a general schema of a possible new approach of the QMS problem.
Our approach is inspired from a view \cite{11,8,12} about the
measurement of classical (non-quantum) random observables. The
general schema from Sec.3 is detailed through a simple
exemplification in Sec.4. We end our considerations in Sec.5 with
some conclusions.

\section{Conjectures and shortcomings}
In their essence the alluded TIUR-connected approaches of QMS
imply someway one or more of the following conjectures
(\textbf{C}):
\begin{itemize}
    \item \textbf{C.1}: The description of QMS must be regarded in an
    indissoluble association with the UR
    \begin{equation}\label{eq:1}
    \Delta A \cdot \Delta B \geqslant \frac{1}{2} \vert \langle
    [\hat{A},\hat{B}]\rangle \vert
    \end{equation}
    supposed as a capital physical formula. Consequently QMS approaches
    must be developed as extensions of TIUR. (Observation: The notations
    in (\ref{eq:1}) are the usual ones from QM and they are shortly reminded
    below in the next section).
    \item \textbf{C.2}: Between quantum and classical measurements there
    is a fundamental distinction due to the exclusive existence of UR
    (\ref{eq:1}) in quantum cases.
    \item \textbf{C.3}: The description of QMS must take into account some
    non-null jumps in the states of the measured system. The respective
    jumps are caused by the perturbative action of measuring devices and
    they are neither avoidable nor negligible.
    \item \textbf{C.4}: A  QMS supplies a single value for a measured
    observable and consequently it must be regarded as a unique (single)
    detection act representable as a collapse (reduction) of the
    corresponding wave function.
\end{itemize}

The mentioned re-examination of TIUR shows \cite{10} that in
reality UR (\ref{eq:1}) are not capital physical formulas.
Consequently we can conclude that it is unreasonable to
subordinate the QMS approaches to the respective UR. But such a
conclusion clearly appears as a true shortcoming for the
conjecture \textbf{C.1}. On the other hand within the same
re-examination one finds that the UR (\ref{eq:1}) belongs to a
general family of fluctuation relations from both quantum and
classical (non-quantum) physics. Then it results that the
mentioned UR cannot motivate a distinction between quantum and
classical situations. Evidently that such a result leaves without
any base the conjecture \textbf{C.2} and it must be noted as a
shortcoming of the respective conjecture.

As regards the conjecture \textbf{C.3} the following facts are
notable. The respective conjecture was not inferred directly from
the main assertions of TIUR. However, it was promoted adjacently
in discussions  generated by TIUR. Firstly, it was said that the
measurements uncertainties are due to the interactions between
measured systems and measuring devices. Secondly, it was added
that the respective interactions cause jumps in the states of the
measured systems. Then it was accredited the supposition that, in
contrast with the classical situations, in QMS the mentioned
uncertainties, interactions and jumps have an unavoidable
character. Subsequently it was promoted the idea that the alluded
measuring jumps must be taken into account in the description of
QMS. In spite of its genesis, the above mentioned idea is proved
to be incorrect by the following genuine and indubitable opinion
\cite{13}: \emph{``it seems essential to the notion of a
measurement that it answers a question about the given situation
existing before measurement. Whether the measurement leaves the
measured system unchanged or brings about a new and different
state of that system is a second and independent question''}. The
natural acceptance of the quoted opinion brings the conjecture
\textbf{C.3} in an insurmountable shortcoming.

The conjecture \textbf{C.4} is contradicted by natural views about
random quantities, from both physics and mathematics. From the
physics viewpoint, the measurement of a random observable
(quantity) must have the same general features, independently of
its quantum or classical nature. However, in the classical context
(e.g. in the study of fluctuation \cite{14,15,11,12}) the
measurement of a macroscopic random observable is not viewed as a
single detection act, associated with some collapse (reduction) of
the corresponding probability distribution. More exactly such a
measurement is regarded \cite{16} as a statistical sampling, i.e.
as an ensemble of great number of individual detection acts. The
respective ensemble gives a nontrivial set of values belonging to
the spectrum of the considered observable. In addition, from a
mathematical viewpoint \cite{17} a random variable (quantity) must
be evaluated not by a unique value but through a statistical set
of values. Then it directly results that because QMS regards
observables with random characteristics they must be viewed as
statistical sampling (in the above-mentioned sense). Consequently
there are no reasons to represent (describe) a QMS as a collapse
(reduction) of a wave function. The mentioned result and
consequence incontestably invalidate the conjecture \textbf{C.4}.
So one finds a shortcoming for the respective conjecture.

The above mentioned shortcomings of the conjectures
\textbf{C.1}-\textbf{4} have an unsurmountable character because
they cannot be combated or avoided by valid arguments derivable
from the TIUR doctrine. But such a fact shows that  TIUR-connected
approaches of QMS are groundless attempts. Then it results that,
at least partially, the problem of QMS description is still an
open question which requires further investigations. In such a
context we think that the new approach that we present in the next
sections can be of interest.

\section{A new approach}
It is known that each approach of QMS description resorts (more or
less explicitly) to some conjectures. Then, for the new approach
aimed here, we suggest the set of the following reconsidered
conjectures (\textbf{RC}):
\begin{itemize}
    \item \textbf{RC.1}: Any measurement searches for  information
    regarding the pre-existent state of the investigated system,
    independently of the quantum or classical nature of the respective
    system.
    \item \textbf{RC.2}: Due to the randomness of quantum observables a
    QMS must consists obligatory in a statistical sampling i.e. in a
    great number of individual detection acts.
    \item \textbf{RC.3}: A description of QMS must contain some extra-QM
    elements regarding the measuring devices and procedures, because
    the mere QM refers only to the intrinsic properties of the
    considered systems.
    \item \textbf{RC.4}: Because of the fact that in the last analysis,
    the results supplied by QMS refer to the measured quantum systems
    they must be evaluated in terms of QM.
\end{itemize}

In mind with \textbf{RC.1}-\textbf{4}, we develop the announced
approach as follows. We consider a spin-less quantum microparticle
with own orbital characteristics described by the intrinsic (IN)
wave function $\Psi_{IN}$. From a theoretical viewpoint
$\Psi_{IN}$ can be regarded as solution of the corresponding
Schr\" odinger equation. In the following probabilistic
considerations, the microparticle is regarded as equivalent with
a statistical ensemble of its own replica taken at the same
instant of time and described by the same wave function
$\Psi_{IN}$. Therefore, for our purposes, the time $t$ appears as
a ``passive'' variable not implied in the randomness of the
considered microparticle. That is why $\Psi_{IN}$ will be written
as a function only of the radius vector $\vec{r}$, i.e.
$\Psi_{IN}=\Psi_{IN}(\vec{r})$. The specific observables
$A_j(j=1,2,\ldots,n)$ of the microparticle are described by the
usual QM operators $\hat{A}_j$(e.g $\hat{x}_{\mu}=x_{\mu}\cdot$
and $\hat{p}_{\mu}=-i\hbar \frac{\partial}{\partial x_{\mu}}
(\mu=1,2,3)$ for Cartesian coordinate and momenta,
$\hat{\vec{p}}=-i\hbar\nabla$ and $\hat{\vec{L}}=-i\hbar
\,\vec{r} \times \nabla$ for momentum and angular momentum
vectors or $\hat{H}=-\frac{\hbar^2}{2m}\nabla^2+V(\vec{r})$ for
Hamiltonian).

Because $A_j$ have random properties, as in probability theory
\cite{17}, for practical purposes they are described by means of
the so-called numerical parameters (or characteristics). In QM the
mostly used such parameters are: the IN-mean-values $\langle
A_j\rangle_{IN}$ the IN-correlations $\mathcal{C}_{IN}(A_j,A_l)$
respectively the IN-standard-deviations $\Delta_{IN}A_j$. Note
that, from a probabilistic perspective, the mentioned numerical
parameters are lower order entities. Additionally, as in
probability theory \cite{17}, the higher order numerical
parameters can also be used (e.g. higher order correlations and
moments). However, such higher order parameters are not usual in
QM literature. As it is known the alluded lower order intrinsic
(IN) numerical parameters are defined by the relations:
\begin{equation}\label{eq:2}
\langle A_{j} \rangle_{IN}=\left(\Psi_{IN},\hat{A}_j
\Psi_{IN}\right) =\int\Psi_{IN}^{*}\left(\vec{r}\right)\hat{A}_{j}
\Psi_{IN}\left( \vec{r} \right) \mathrm{d}^{3}\vec{r}
\end{equation}
\begin{equation}\label{eq:3}
\mathcal{C}_{IN}(A_{j},A_{l})=\left(\delta_{IN}\hat{A}_{j}\Psi_{IN},
\delta_{IN}\hat{A}_{l}\Psi_{IN}\right)\;, \quad
\delta_{IN}\hat{A}_{j}=\hat{A}_{j}-\langle A_{j}\rangle_{IN}
\end{equation}
\begin{equation}\label{eq:4}
\Delta_{IN}A_{j}=\sqrt{\mathcal{C}_{IN}(A_{j},A_{j})}
\end{equation}
In (\ref{eq:2}) and (\ref{eq:3}) $(f_a,f_b)$ denotes the scalar
product of the functions $f_a$ and $f_b$.

From a general physical perspective the intrinsic parameters
(\ref{eq:2})-(\ref{eq:4}) must be compared with the corresponding
recorded (or ascertained) parameters considered as being given by
measurements. But if in connection with the measurements, besides
the practical experimental actions, one wants to operate with
theoretical descriptions the term ``recorded'' must be regarded in
two postures. In one posture it has a significance of ``factual
records'' (FR) and refers to the data supplied by adequate
practical experiments. In other posture the respective term has a
significance of ``prognosticated records'' (PR) and refers to the
quantities predicted by the considered theoretical description.

For the observables $A_j(j=1,2,\ldots,s)$ the alluded
FR-parameters can be defined as follows. Because $A_j$ are random
variables their measurements must be regarded as statistical
samplings (done on statistical replies of a system (microparticle)
considered in the same pre-measurement state). By such a sampling,
on the recorder of the measuring device, for each observable $A_j$
one obtains a set of experimental values noted as:
$a_{j1},a_{j2},\ldots,a_{jn}$. Then the ensemble of observables $A_j$
can be characterized by means of the following FR-parameters
(defined according to the mathematical statistics rules
\cite{17}).
\begin{equation}\label{eq:5}
    \langle A_j \rangle_{FR} =\frac{1}{n} \sum_{k=1}^{n} a_{jk}
\end{equation}
\begin{equation}\label{eq:6}
    \mathcal{C}_{FR} (A_j,A_l)=\frac{1}{n}\sum_{k=1}^{n} (a_{jk}
    -\langle A_j\rangle_{FR})(a_{lk}-\langle A_l\rangle_{FR})
\end{equation}
\begin{equation}\label{eq:7}
    \Delta_{FR}A_j=\sqrt{\mathcal{C}_{FR}(A_j,A_j)}
\end{equation}
For the parameters (\ref{eq:5})-(\ref{eq:7}) can be used the
denominations FR-mean-value, FR-correlation respectively
FR-standard-deviation.

Evidently that the FR-quantities (\ref{eq:5})-(\ref{eq:7}) depend
both on the intrinsic properties of the measured system and on the
characteristics of the measuring devices/procedures. That is why
the mentioned FR-quantities are significant only if they are
considered in connection with the experimental setting which
supply the values $a_{j1},a_{j2},\ldots,a_{jn}$.

On the other hand if one wishes to operate with a theoretical
description of QMS the IN-parameters (\ref{eq:2})-(\ref{eq:4})
must be compared with corresponding parameters of ``prognosticated
records'' (PR) - type from an adequate mathematical model. For
such a model we consider that the respective PR-parameters are
defined similarly with (\ref{eq:2})-(\ref{eq:4}) by means of a
PR-wave-function $\Psi_{PR}$ and with the same operators, i.e.
\begin{equation}\label{eq:8}
\langle A_{j} \rangle_{PR}=\left(\Psi_{PR},\, \hat{A}_j
\Psi_{PR}\right) = \int \Psi_{PR}^{*}\left(
\vec{r}\right)\hat{A}_{j} \Psi_{PR}\left( \vec{r} \right)
\mathrm{d}^{3}\vec{r}
\end{equation}
\begin{equation}\label{eq:9}
\mathcal{C}_{PR}(A_{j},A_{l})=\left(\delta_{PR}\hat{A}_{j}\Psi_{PR}
, \, \delta_{PR}\hat{A}_{l}\Psi_{PR}\right)\;, \quad
\delta_{PR}\hat{A}_{j}=\hat{A}_{j}-\langle A_{j}\rangle_{PR}
\end{equation}
\begin{equation}\label{eq:10}
\Delta_{PR}A_{j}=\sqrt{\mathcal{C}_{PR}(A_{j},A_{j})}
\end{equation}
Our above consideration is motivated by the known fact that, in
theoretical descriptions, the randomness of a quantum
microparticle is incorporated in its wave function but not in
operators of its observables. Properties of various states of a
microparticle are described by different wave functions but with
the same operators. A similar situation exists in the case of
classical statistical systems for which the randomness is
incorporated in the probability densities but not in the
expressions of macroscopic random variables. In the alluded cases
also the properties of various states of a system are described
with different probability densities but with the same expressions
for the macroscopic random variables. In a classical case a
measurement is described similarly \cite{11,12} by appealing to a
``recorded'' density of probability. The term ``recorded'' from
\cite{8,9,10,11,12} implies the same significance as the here used
term ``prognosticated records''. Note that in both quantum and
classical cases the appeals to ``prognosticated records'' or
``recorded'' entities (wave function or probability density) must
not be regarded as a description of collapse (reduction) for the
corresponding intrinsic entities.

By adopting the relations (\ref{eq:8})-(\ref{eq:10}) the task of
our approach becomes to express $\Psi_{PR}$ (or related
quantities) in terms of $\Psi_{IN}$ (or associated  entities) and
of some elements regarding the measuring devices. For such a
task, firstly we show that the parameters
(\ref{eq:2})-(\ref{eq:4}) and (\ref{eq:8})-(\ref{eq:10}) can be
expressed in terms of certain quantities connected with
$\Psi_{Y}(Y=IN;PR)$ and having ordinary probabilistic
significance in the sense of probability theory \cite{17}. So we
transcribe $\Psi_{Y}$ in the form $\Psi_{Y}=\vert\Psi_{Y}\vert
\exp(i\Phi_Y)$ where $\vert\Psi_{Y}\vert$ and $\Phi_Y$ denote the
modulus respectively the argument of $\Psi_Y$. As quantities
of the mentioned type we take firstly the probability densities associated with
$\Psi_Y$ and defined by
\begin{equation}\label{eq:11}
\rho_{Y}=|\Psi_{Y}|^{2}
\end{equation}
Other quantities with ordinary probabilistic significance are the
probability currents (or probability fluxes per unit-area):
\begin{equation}\label{eq:12}
\vec{J}_{Y}=-\frac{i\hbar}{2m}\left( \Psi_{Y}^{*}\nabla\Psi_{Y} -
\Psi_{Y}\nabla\Psi_{Y}^{*} \right)=\frac{\hbar}{m}|\Psi_{Y}|^{2}
\cdot\nabla\Phi_{Y}
\end{equation}
($m$ denotes the mass of microparticle).

Now let us show that the parameters (\ref{eq:2})-(\ref{eq:4}) and
(\ref{eq:8})-(\ref{eq:10}) can be expressed in terms of $\rho_{Y}$
and $\vec{J}_{Y}$. Then we observe that if an operator $\hat{A}$
does not depend on $\nabla$, i.e. $\hat{A}=\hat{A}(\vec{r})$ in
(\ref{eq:2}) and (\ref{eq:8}) can be used the substitutions:
\begin{equation}\label{eq:13}
\Psi_{Y}^{*}\hat{A}\Psi_{Y}=A(\vec{r})\,\rho_{Y}
\end{equation}
On the other hand if $\hat{A}$ depends on $\nabla$, i.e. $\hat{A}=
\hat{A}(\nabla)$, by taking $\Psi_{Y}=|\Psi_{Y}|\exp (i \Phi_{Y})$
and using (\ref{eq:11})-(\ref{eq:12}) in (\ref{eq:2}) and
(\ref{eq:8}) one can resort to the substitutions like:
\begin{equation}\label{eq:14}
\Psi_{Y}^{*}\nabla\Psi_{Y}=\frac{1}{2}\nabla\rho_{Y}+
\frac{im}{\hbar}\vec{J}_{Y}
\end{equation}
\begin{equation}\label{eq:15}
\Psi_{Y}^{*}\nabla^{2}\Psi_{Y}=\rho_{Y}^{1/2}\nabla^{2}\rho_{Y}^{1/2}
+\frac{im}{\hbar}\nabla\vec{J}_{Y}-\frac{m^2}{\hbar^2}
\frac{\vec{J_Y}^2}{\rho_{Y}}
\end{equation}

The existence of substitutions (\ref{eq:13})-(\ref{eq:15})
suggests that the description of QMS can be completed by adequate
considerations about the quantities $\rho_{Y}$ and $\vec{J}_{Y}$.
As the respective quantities have ordinary probabilistic
significance for the alluded completion we resort to the model
used \cite{11,12} in the description of measurements of classical
random observables. We also take into account the fact that
$\rho_{Y}$ and $\vec{J}_{Y}$ refer to the positional respectively
motional aspects of probabilities. Or, from an experimental
perspective, the two aspects can be regarded as measurable by
independent devices and procedures. Then the alluded completion
must consist in giving independent relationships between
$\rho_{PR}$ and $\rho_{IN}$ on the one hand respectively between
$\vec{J}_{PR}$ and $\vec{J}_{IN}$ on the other hand. The mentioned
relationships can be expressed  formally by the following generic
formulas:
\begin{equation}\label{eq:16}
\rho_{PR}=\hat{G}\rho_{IN}
\end{equation}
\begin{equation}\label{eq:17}
J_{PR;\mu}=\sum_{\nu=1}^{3}\hat{\Lambda}_{\mu\nu}J_{IN;\nu}
\end{equation}
($J_{Y;\mu}$ with $Y=IN,PR$ and $\mu=1,2,3=x,y,z$ denote the
Cartesian components of  vectors $\vec{J}_{Y}$). In (\ref{eq:16})
and (\ref{eq:17}) $\hat{G}$ and $\hat{\Lambda}_{\mu,\nu}$ signify
the measurements operators. They must comprise obligatory
characteristics of measuring devices and procedures. So $\hat{G}$
and $\hat{\Lambda}_{\mu,\nu}$ must contain some extra-QM elements,
i.e. elements that do not belong to the usual QM description of
the intrinsic properties of the measured microparticles.

For measuring devices with linear and stationary characteristics,
similarly with the classical case \cite{11,12}, the relations
(\ref{eq:16})-(\ref{eq:17}) can be written as:
\begin{equation}\label{eq:18}
\rho_{PR}(\vec{r})=\int G(\vec{r},\vec{r}\,')\, \rho_{IN}(
\vec{r}\,')\,\mathrm{d}^{3}\vec{r}\,'
\end{equation}
\begin{equation}\label{eq:19}
J_{PR;\mu}(\vec{r})=\sum_{\nu=1}^{3}\int \Lambda_{\mu\nu}
(\vec{r},\vec{r}\,')\,J_{IN;\nu}(\vec{r}\,')\,\mathrm{d}^{3}\vec{r}\,'
\end{equation}
The kernels $G(\vec{r},\vec{r}\,')$ and
$\Lambda_{\mu\nu}(\vec{r},\vec{r}\,')$ are supposed to satisfy the
conditions:
\begin{equation}\label{eq:20}
\int G(\vec{r},\vec{r}\,')\,\mathrm{d}^{3}\vec{r}= \int
G(\vec{r},\vec{r}\,')\, \mathrm{d}^{3}\vec{r}\,'=1
\end{equation}
\begin{equation}\label{eq:21}
\int \Lambda_{\mu\nu}(\vec{r},\vec{r}\,')\,\mathrm{d}^{3}\vec{r} =
\int\Lambda_{\mu\nu}(\vec{r},\vec{r}\,')\,\mathrm{d}^{3}\vec{r}\,'=1
\end{equation}
These conditions show the one-to-one probabilistic correspondence
between the intrinsic quantities $\rho_{IN}$ and $\vec{J}_{IN}$
respectively the recorded ones $\rho_{PR}$ and $\vec{J}_{PR}$.
Parameters (\ref{eq:8})-(\ref{eq:10}), evaluated by means of the
relations (\ref{eq:13})-(\ref{eq:15})and
(\ref{eq:16})-(\ref{eq:19}), incorporate randomness of both
intrinsic and extrinsic nature, corresponding to the own
properties of the investigated microparticle respectively to the
measuring devices. As evaluated the mentioned parameters have a
theoretical significance. Their adequacy must be tested by
comparing with the corresponding FR-parameters
(\ref{eq:5})-(\ref{eq:7}) obtained by statistical processing of
the real experimental data. If the test is affirmative both
descriptions, of intrinsic QM properties respectively of QMS, can
be accepted as adequate. However, if the test invalidates the
theoretical results, at least one of the respective descriptions
must be regarded as  inadequate.

From the origins of their history, the QMS approaches are
concerned with the problem of quantitative evaluation for
measuring uncertainties (i.e. for errors induced by the
measurements in the values of the measured quantum observables).
That is why it is of interest to discuss the respective problem in
connection with the here promoted approach. Our discussion starts
by pointing out the fact that quantum observables have a random
character. Consequently, the uncertainties of such an observable
must be evaluated through   indicators, which comprise information
from the whole its spectrum. It is easy to see that indicators of
the alluded kind can be introduced by means of the numerical
parameters defined by relations (\ref{eq:5})-(\ref{eq:7}) and
(\ref{eq:8})-(\ref{eq:10}). That is why we suggest that,
conjointly with the above-presented approach of QMS, the measuring
uncertainties to be evaluated through the following uncertainty
(or error) indicators of FR-type respectively PR-type:
\begin{equation}\label{eq:22}
\delta_{FR}\left( \langle A_{j} \rangle \right)=\left| \langle
A_{j} \rangle _{FR} - \langle A_{j} \rangle _{IN} \right|
\end{equation}
\begin{equation}\label{eq:23}
\delta_{FR} \left( \mathcal{C}(A_{j},A_{l}) \right)= \left|
\mathcal{C}_{FR}(A_{j},A_{l})-\mathcal{C}_{IN}(A_{j},A_{l})\right|
\end{equation}
\begin{equation}\label{eq:24}
\delta_{FR} (\Delta A_{j})=\left|\Delta_{FR}A_{j}-
\Delta_{IN}A_{j}\right|
\end{equation}

\begin{equation}\label{eq:25}
\delta_{PR}\left( \langle A_{j} \rangle \right)=\left| \langle
A_{j} \rangle _{PR} - \langle A_{j} \rangle _{IN} \right|
\end{equation}
\begin{equation}\label{eq:26}
\delta_{PR} \left( \mathcal{C}(A_{j},A_{l}) \right)= \left|
\mathcal{C}_{PR}(A_{j},A_{l})-\mathcal{C}_{IN}(A_{j},A_{l})\right|
\end{equation}
\begin{equation}\label{eq:27}
\delta_{PR} (\Delta A_{j})=\left|\Delta_{PR}A_{j}-
\Delta_{IN}A_{j}\right|
\end{equation}

The above defined uncertainty indicators of FR-type
(\ref{eq:22})-(\ref{eq:24}) respectively of PR-type
(\ref{eq:25})-(\ref{eq:27}) are significant for a given practical
measurement respectively for a considered theoretical description
of QMS. The concordance degree between the two types of
indicators shows the level of adequation of the theoretical
description in respect with the considered measurement.

The uncertainty indicators of PR-type (\ref{eq:25})-(\ref{eq:27})
have a restricted significance for a system (microparticle),
because they refer to some particular observables of the
respective system. A more generic uncertainty indicators, also of
PR-type but  regarding a system in the whole, can be introduced
by means of the following informational entropies of Shanon type:
\begin{equation}\label{eq:28}
\mathcal{H}_{Y}=-\int \rho_{Y} \ln \rho_{Y}\,
\mathrm{d}^{3}\vec{r}
\end{equation}
\begin{equation}\label{eq:29}
\tau_{Y}=-\int |\vec{J}_{Y}| \ln |\vec{J}_{Y}|\,
\mathrm{d}^{3}\vec{r}
\end{equation}
Here $\mathcal{H}_{Y}$ and $\tau_{Y}$ can be called positional
respectively motional informational entropies. Then the alluded
generic uncertainty indicators can be defined as
\begin{equation}\label{eq:30}
\delta_{PR} \mathcal{H}=\mathcal{H}_{PR}-\mathcal{H}_{IN}
\end{equation}
\begin{equation}\label{eq:31}
\delta_{PR}\tau=\tau_{PR}-\tau_{IN}
\end{equation}
It is interesting to note the fact that within the above-presented
description of QMS the indicator $\delta_{PR}\mathcal{H}$ is a
nonnegative quantity (i.e. $\delta_{PR}\mathcal{H}\geqslant 0$).
The respective fact can be proved, similarly with the classical
situation \cite{11,12}, by means of the relations (\ref{eq:18})
and (\ref{eq:20}). So by taking into account the respective
relations, the normalization of both $\rho_{IN}$ and $\rho_{PR}$,
and the evident formula $\ln y \leq y-1\; (y>0)$ one can write:
\begin{eqnarray}\label{eq:32}
\delta_{PR}\mathcal{H}&=&\mathcal{H}_{PR}-\mathcal{H}_{IN}=\nonumber \\
&=&-\int \mathrm{d}^{3}\vec{r} \int \mathrm{d}^{3}\vec{r}\,'\,
G(\vec{r},\vec{r}\,')\, \rho_{IN}(\vec{r}\,')^{2} \ln
\frac{\rho_{PR}(\vec{r})}{\rho_{IN}(\vec{r}\,')} \geqslant \nonumber \\
&\geqslant& -\int \mathrm{d}^{3}\vec{r} \int
\mathrm{d}^{3}\vec{r}\,'\, G(\vec{r},\vec{r}\,')\,
\rho_{IN}(\vec{r}\,') \left[
\frac{\rho_{PR}(\vec{r})}{\rho_{IN}(\vec{r}\,')} -1 \right]=0
\end{eqnarray}

The above considerations give a genuine description of QMS in
which one finds, in adequate positions, all the essential
elements. The respective elements include: (i) the intrinsic
numerical parameters (\ref{eq:2})-(\ref{eq:4}), (ii) the model
represented by (\ref{eq:16})- (\ref{eq:21}) for describing the
influences of measuring devices, (iii) the recorded numerical
parameters (\ref{eq:8})-(\ref{eq:10}) and (iv) the uncertainties
indicators of PR-type (\ref{eq:25})-(\ref{eq:27}) or
(\ref{eq:30})-(\ref{eq:31}).

In the end of this section we note  that the description of QMS
presented here, as well as the one discussed in \cite{11,12} for
classical measurements, can be regarded formally from the
perspective of information theory. In such a perspective, a
measurement appears as a process of information transmission. The
source of information is the measured system and the intrinsic
values of its observables represent the input information. The
chain of measuring devices plays the role of channel for
information transmission. The recorded data about the measured
observables represent the output information. Then the
measurement uncertainties can be regarded as alterations of the
transmitted information.

\section{A simple exemplification}
To illustrate the above-introduced QMS approach let us refer to
the following simple exemplification. We consider a quantum
microparticle in a one-dimensional motion along the $x$-axis. Its
own properties are supposed to be described by the intrinsic wave
function $\Psi_{IN}(x) =|\Psi_{IN}(x)|\exp\left( i\,\Phi_{IN}(x)
\right)$ with:
\begin{equation}\label{eq:33}
\vert\Psi_{IN}(x)\vert=\left( \alpha \sqrt{2\pi}
\right)^{-1/2}\exp\left\{ -\frac{(x-x_{0})^{2}}{4\alpha^{2}}
\right\}\,, \quad \Phi(x)=kx
\end{equation}
Then the intrinsic probability density and current defined by
(\ref{eq:11}) and (\ref{eq:12}) are:
\begin{equation}\label{eq:34}
\rho_{IN}(x)=\frac{1}{\alpha\sqrt{2\pi}}\,
\exp{\left\{{-\frac{(x-x_{0})^{2}}{2\alpha^{2}}}\right\}}
\end{equation}
\begin{equation}\label{eq:35}
J_{IN}(x)=\frac{\hbar k}{m\alpha\sqrt{2\pi}}
\exp{\left\{{-\frac{(x-x_{0})^{2}}{2\alpha^{2}}}\right\}}
\end{equation}
So the intrinsic characteristics of the microparticle are
described by the parameters $x_{0}$, $\alpha$ and $k$.

Considering that the errors of QMS are small in (\ref{eq:18}) and
(\ref{eq:19}), one can operate with the one-dimensional kernels
of Gaussian forms given by:
\begin{equation}\label{eq:36}
G(x,x')=\frac{1}{\sigma \sqrt{2\pi}} \exp{\left\{
-\frac{(x-x')^{2}}{2\sigma^{2}} \right\}}
\end{equation}
\begin{equation}\label{eq:37}
\Lambda(x,x')=\frac{1}{\lambda\sqrt{2\pi}} \exp{\left\{
-\frac{(x-x')^{2}}{2\lambda^{2}} \right\}}
\end{equation}
Here $\sigma$ and $\lambda$ describe the error characteristics of
the measuring devices (see bellow).

By using (\ref{eq:36})-(\ref{eq:37}) in the one-dimensional
versions of the relations (\ref{eq:18})-(\ref{eq:19}) one finds:
\begin{equation}\label{eq:38}
\rho_{PR}(x)= \frac{1}{\sqrt{2\pi(\alpha^{2}+\sigma^{2})}}\,
\exp{\left\{ -\frac{(x-x_{0})^{2}}{2(\alpha^{2}+\sigma^{2})}
\right\}}
\end{equation}
\begin{equation}\label{eq:39}
J_{PR}(x)=\frac{\hbar k}{m\sqrt{2\pi(\alpha^{2}+\lambda^{2})}}\,
\exp{\left\{ -\frac{(x-x_{0})^{2}}{2(\alpha^{2}+\lambda^{2})}
\right\}}
\end{equation}

One can see that in the case when \mbox{$\sigma\to 0$} and
\mbox{$\lambda \to 0$} the kernels $G(x,x')$ and $\Lambda(x,x')$
degenerate into the Dirac function $\delta(x-x')$. Then
\mbox{$\rho_{PR}(x)\rightarrow \rho_{IN}(x)$} and
$J_{PR}(x)\rightarrow J_{IN}(x)$. Such a case corresponds to an
ideal measurement. Alternatively the cases with $\sigma\neq 0$
and/or $\lambda\neq 0$ are associated with non-ideal measurements.

As observables of interest, we consider the coordinate $x$ and
momentum $p$ described by the operators $\hat{x}=x\cdot$ and
$\hat{p}= -i\hbar \frac{\partial}{\partial x}\,$. Adequately we use
the expressions (\ref{eq:34})-(\ref{eq:35}) and
(\ref{eq:38})-(\ref{eq:39}) in the relations
(\ref{eq:2})-(\ref{eq:4}) and (\ref{eq:8})-(\ref{eq:10}). Then,
by using (\ref{eq:13})-(\ref{eq:15}), for the mentioned
observables one finds the following numerical parameters of
IN-type respectively of PR-type.
\begin{equation}\label{eq:40}
\langle x \rangle_{IN}= \langle x \rangle_{PR}=x_{0}\;,  \qquad
\langle p \rangle_{IN}= \langle p \rangle_{PR}=\hbar k
\end{equation}
\begin{equation}\label{eq:41}
\mathcal{C}_{IN}(x,p)=\mathcal{C}_{PR}(x,p)=\frac{i\hbar}{2}
\end{equation}
\begin{equation}\label{eq:42}
\Delta_{IN}x=\alpha\;, \qquad
\Delta_{PR}x=\sqrt{\alpha^{2}+\sigma^{2}}
\end{equation}
\begin{equation}\label{eq:43}
\Delta_{IN}p=\frac{\hbar}{2\alpha}
\end{equation}
\begin{equation}\label{eq:44}
\Delta_{PR}p=\hbar \sqrt{\frac{k^{2}(\alpha^{2}+\sigma^{2})} {\sqrt{\alpha^{4}-\lambda^{4}+
2\sigma^{2}(\alpha^{2}+\lambda^{2})}}
-k^{2}+\frac{1}{4(\alpha^{2}+\sigma^{2})}}
\end{equation}
Then for the considered observables $x$ and $p$ the uncertainty (error)
indicators of PR-type (\ref{eq:25})-(\ref{eq:27}) become:
\begin{equation}\label{eq:45}
\delta_{PR}\left( \langle x \rangle \right)=0 \,, \quad
\delta_{PR}\left( \langle p \rangle \right)=0 \,, \quad
\delta_{PR}\left( \mathcal{C} (x,p) \right)=0
\end{equation}
\begin{equation}\label{eq:46}
\delta_{PR} \left( \Delta x \right) =
\sqrt{\alpha^{2}+\sigma^{2}}-\alpha
\end{equation}
\begin{equation}\label{eq:47}
\fl\delta_{PR}\left( \Delta p \right)=\hbar \left\{
\sqrt{\frac{k^{2}(\alpha^{2}+\sigma^{2})} {\sqrt{\alpha^{4}-
\lambda^{4}+2\sigma^{2}(\alpha^{2}+\lambda^{2})}} -k^{2}+\frac{1}{4(\alpha^{2}+\sigma^{2})}}
-\frac{1}{2\alpha}\right\}
\end{equation}
These relations show that for the considered association
microparticle-QMS the numerical parameters $\langle x\rangle$,
$\langle p\rangle$ and $\mathcal{C}(x,p)$ are not affected  by
uncertainties (errors). However, for the same association the parameters $\Delta
x$ and $\Delta p$ are troubled by the measurement, the
corresponding non-null uncertainty (error) indicators of PR-type
being given by (\ref{eq:46})-(\ref{eq:47}).

Now, for the here discussed model of QMS description, let us
search the entropic error indicators of PR-type defined by the
relations (\ref{eq:28})-(\ref{eq:31}). By using the expressions
(\ref{eq:33})-(\ref{eq:35}) and (\ref{eq:38})-(\ref{eq:39}) one finds:
\begin{equation}\label{eq:48}
\delta_{PR}\mathcal{H}=\frac{1}{2}\ln{\left(
1+\frac{\sigma^{2}}{\alpha^{2}} \right)}
\end{equation}
\begin{equation}\label{eq:49}
\delta_{PR}\tau =\frac{\hbar k}{2m}\ln{\left( 1+
\frac{\lambda^{2}}{\alpha^{2}}\right)}
\end{equation}
If in (\ref{eq:33}) we restrict to the values $x_{0}=0$, $k=0$ and
$\alpha=\sqrt{\frac{\hbar}{2m\omega}}$, our system is just a
quantum oscillator with mass $m$ and pulsation $\omega$ situated
in its ground state. The corresponding numerical parameters and
error indicators for observables $x$ and $p$ can be obtained from
(\ref{eq:40})-(\ref{eq:44}) respectively
(\ref{eq:45})-(\ref{eq:47}) by means of the mentioned
restrictions. However, in the case of oscillator it is
interesting to point out the measuring characteristics for another
observable, namely for energy described by the Hamiltonian:
\begin{equation}\label{eq:50}
\hat{H}=\frac{1}{2m}\hat{p}^2+\frac{m\omega^2}{2}\hat{x}^2
\end{equation}
Then, for the probabilistic numerical parameters of oscillator energy one
finds:
\begin{equation}\label{eq:51}
\langle H\rangle_{IN}=\frac{\hbar\omega}{2}\;, \qquad
\Delta_{IN}H=0
\end{equation}
\begin{equation}\label{eq:52}
\langle H\rangle_{PR}=\frac{\omega \left[ \hbar^2+\left(\hbar
+2m\omega \sigma^2\right)^2 \right]}{4(\hbar+2m\omega\sigma^2)}
\end{equation}
\begin{equation}\label{eq:53}
\Delta_{PR}H=\frac{\sqrt{2}m\omega^{2}\sigma^{2}(\hbar+m\omega\sigma^{2})}
{(\hbar+2m\omega\sigma^{2})}
\end{equation}
The corresponding PR-uncertainty (error) indicators are:
\begin{equation}\label{eq:54}
\delta_{PR}\left(\langle H\rangle\right)=\frac{\omega \left[
\hbar^2+ \left(\hbar+2m\omega \sigma^2\right)^2
\right]}{4(\hbar+2m\omega\sigma^2)}-\frac{\hbar\omega}{2}
\end{equation}
\begin{equation}\label{eq:55}
\delta_{PR}(\Delta H)=\frac{\sqrt{2}m\omega^{2}\sigma^{2}
(\hbar+m\omega\sigma^{2})}{(\hbar+2m\omega\sigma^{2})}
\end{equation}

\section{Conclusions}
The problem of QMS description persists in our days as an open
and disputed question. Many of its approaches are TIUR-connected
because they are founded on conjectures mainly inspired from
TIHR. But indubitable facts \cite{10} show that TIUR is an
incorrect doctrine. Consequently all arguments founded on TIUR
imply important deficiencies. Particularly in the case of
TIUR-connected approaches of QMS the main conjectures are
affected by insurmountable shortcomings. Such a finding motivates
our interest for a possible new approach of QMS, based on a set
of reconsidered and natural conjectures. We propose a set of four
such conjectures and develop an adequate approach of QMS.

Our approach is founded on the usual probabilistic conception of
QM. Therefore, for a quantum microparticle we operate  with
probabilistic entities (probability density and probability
current) respectively with QM operators. We opine that, because
in practice a correct QMS must consist in a statistical sampling,
from a theoretical viewpoint a QMS must be represented as
processing of the mentioned probabilistic entities while the
quantum operators remain unchanged. Similarly, with the
description of classical (non-quantum) measurements, the alluded
processing must be pictured as changes of the respective
entities. We opine that for a wide class of situations such
changes can be modeled  as linear integral transforms. Therefore,
both probability density and probability current appear in
intrinsic respectively ``prognosticated records'' posture. In the
first posture, they regard the own characteristics of the
measured microparticle, while in the second posture they comprise
information related both to the respective microparticle and to
the measuring devices. The information regarding the measuring
devices is introduced theoretically through the adopted model in
description of QMS.

Together with the mentioned features of QMS the quantum
observables must be naturally evaluated through the probabilistic numerical
parameters such as mean values, correlations and standard
deviations. Within the discussed approach, the respective
parameters are characterized by intrinsic (IN) respectively
``prognosticated records'' (PR) values. Such values are calculable
by means of QM operators but with  IN- respectively PR-
expressions for probability density and probability current. Then
a natural description of measuring uncertainties for quantum
observables is expressible in terms of differences between the
mentioned ``prognosticated records'' and intrinsic values. Another
description of measurements uncertainties, more generic (i.e. not
associated with some particular observables), can be done in
terms of informational entropies of Shanon type.

The here recapitulated features of our QMS approach are detailed
from a general perspective in Sec.3, while in Sec.4 they are
illustrated by means of a simple exemplification.

We remind here that our QMS approach is quite different from the
TIUR-connected approaches (founded on (or inspired from) TIHR).
The difference is evidenced on the one hand by the idea that QMS
must be regarded as statistical samplings but not as individual
detection acts. Consequently,  we can avoid completely the
controversial conception of wave function collapse (reduction).
On the other hand, the alluded difference is pointed out by our
presumption that the description of QMS must be regarded as a
distinct and independent task comparatively with the usual QM
procedures. Accordingly, with the respective regards the
description of measurement must be considered and discussed as a
scientific branch self-determined and additional comparatively
with the quantum or classical chapters of physics. The mentioned
chapters, as in fact is well established by the scientific
practice, investigate only the intrinsic properties of the
physical systems.

\ack

The above-presented ideas about QMS were announced preliminarily
in the e-print \cite{18}. Now they are reported here in a somewhat
revised form in order to offer a possible completion for our
opinions about TIUR, reiterared and consolidated in a more recent
text \cite{10}.

I wish to express my deep gratitude to the World Scientific
Publishing Company for putting at my disposal a copy of the
monumental book \cite{1}.

The work reported in the present text benefited partially of some
facilities offered by grants from the Romanian Ministry of
Education and Research.

\section*{List of abbreviations}
\begin{tabular}{ll}
\textbf{C} & = conjectures  \\
FR & = ``factual records''  \\
IN & = intrinsic  \\
PR & = ``prognosticated records''   \\
QM & = quantum mechanics   \\
QMS & = quantum measurements  \\
\textbf{RC} & = reconsidered conjectures  \\
TIUR & = traditional interpretation of uncertainty relations  \\
UR & = uncertainty relations
\end{tabular}


\section*{References}
\begin{thebibliography}{99}
\bibitem{1} Auletta G 2000 \textit{Foundations and Interpretation of Quantum
Mechanics} (World Scientific, Singapore)
\bibitem{2} Lists \textit{New Preprints and Reports in the CERN
Library} http://weblib.cern.ch/
\bibitem{3} arXiv.org e-Print archive - Quantum Physics:
http://mentor.lanl.gov/archive/quant-ph
\bibitem{4} Dumitru S 1977 \textit{Epistemological Letters} \textbf{15} 1
\bibitem{5} Dumitru S 1980 \textit{CERN Library Preprint} PRE-24165
\bibitem{6} Dumitru S 1984 \textit{Microphysics (Solved Problems and a
Critical Analysis of the Question of Significance of the
Uncertainty Relations)} (Cluj-Napoca: Dacia) (in Roumanian)
\bibitem{7} Dumitru S 1987 in \textit{Recent Advances in Statistical
Physics} Ed. Datta B and Dutta M (Singapore: World Scientific)
\bibitem{8} Dumitru S 1988 \textit{Rev.Roum.Phys.} \textbf{33} 11
\bibitem{9} Dumitru S 2000 \textit{Preprint} arXiv quant-ph/0004013
\bibitem{10} Dumitru S 2002 \textit{Preprint} arXiv quant-ph/0206009
\bibitem{11} Dumitru S 1974 \textit{Phys. Lett. A} \textbf{48} 109
\bibitem{12} Dumitru S 1999 \textit{Optik (Stuttgart)} \textbf{110} 110
\bibitem{13} Albertson J 1963 \textit{Phys. Rev.} \textbf{129} 940
\bibitem{14} Landau L Lifchitz E 1984 \textit{Physique Statistique}
(Mir: Moscou)
\bibitem{15} Diu B Guthmann C Lederer D  and Roulet B 1995
\textit{Elements de Physique Statistique} (Hermann: Paris)
\bibitem{16} Shilling H l972 \textit{Statistiche Physik in Beispielen}
(Veb Fachbucherverlag, Leipzig)
\bibitem{17} Korn G A Korn T M 1968 \textit{Mathematical Handbook}
(Mc Graw Hill, New York) (Russian version 1977 (Nauka: Moscow))
\bibitem{18} Dumitru S 2001 \textit{Preprint} arXiv quant-ph/0111143
\end{thebibliography}
\end{document}